\documentclass[12pt]{article}
\usepackage{amsfonts,amsmath}

\usepackage{color}
\usepackage{tabularx}

\usepackage{dsfont}
\usepackage{times}

\usepackage[dvips]{graphicx}
\usepackage{ifthen}
\usepackage{enumerate}
\usepackage{amsmath}
\usepackage{amssymb}
\usepackage[mathscr]{eucal}
\usepackage[errorshow]{tracefnt}
\usepackage{amsmath}
\usepackage{slashed}
\usepackage{amssymb}
\usepackage{array}
\usepackage{amsthm}
\usepackage{eucal}
\usepackage{eufrak}
\usepackage{epsf}
\usepackage{epsfig}
\usepackage{psfrag}
\usepackage{multirow}
\usepackage[apple mac]{inputenc}
\def\+{{+\!\!\!+}}

\def\calb         {{\cal B}}
\def\calc         {{\cal C}}
\def\cald         {{\cal D}}

\def\calF         {{\cal F}}

\def\call         {{\cal L}}

\def\caln         {{\cal N}}

\def\pmb#1{\setbox0=\hbox{#1}% 
\kern.0em\copy0\kern-\wd0 
\kern-.04em\copy0\kern-\wd0 
\kern.08em\copy0\kern-\wd0 
\kern-.04em\raise.0433em\box0 }         %poor man's bold macro (TexBook) 
\newcommand{\nc}{\newcommand} 
\nc{\beq}{\begin{equation}} 
\nc{\eeq}[1]{\label{#1}\end{equation}} 
\nc{\ber}{\begin{eqnarray}} 
\nc{\eer}[1]{\label{#1}\end{eqnarray}} 
\nc{\pek}[1]{\cite{#1}} 
\nc{\enr}[1]{(\ref{#1})} 
\nc{\kal}[1]{{\cal{#1}}} 
\nc{\dott}{\;\cdot\;} 
\def\0 {\nonumber}

\begin{document} 

%\maketitle
\setcounter{page}{0}
\newcommand{\inv}[1]{{#1}^{-1}} %inverse 
\renewcommand{\theequation}{\thesection.\arabic{equation}} 
\newcommand{\be}{\begin{equation}} 
\newcommand{\ee}{\end{equation}} 
\newcommand{\bea}{\begin{eqnarray}} 
\newcommand{\eea}{\end{eqnarray}} 
\newcommand{\re}[1]{(\ref{#1})} 
\newcommand{\qv}{\quad ,} 
\newcommand{\qp}{\quad .} 

\def\qp{Q_+}
\def\qm{Q_-}
\def\qbp{\bar Q_+}
\def\qbm{\bar Q_-}
\def\sgh{\Sigma_{g,h}}

\begin{titlepage}
%\title{}
\begin{center}

\hfill SISSA 17/2010/FM-EP\\
%\hfill   hep-th/yymmnnn\\

\vskip .3in \noindent

%\vskip .1in

{\Large \bf{Taming open/closed string duality with a Losev trick}} \\

\vskip .2in

{\bf Giulio Bonelli, Andrea Prudenziati and Alessandro Tanzini}

\vskip .05in
{\em\small International School of Advanced Studies (SISSA) \\ and \\ INFN, Sezione di Trieste \\
 via Beirut 2-4, 34014 Trieste, Italy}
\vskip .5in
\end{center}
\begin{center} {\bf ABSTRACT }
\end{center}
\begin{quotation}\noindent
A target space string field theory formulation for open and closed B-model is provided
by giving a Batalin-Vilkovisky quantization of the holomorphic Chern-Simons theory with off-shell gravity background.
The target space expression for the coefficients of the holomorphic anomaly equation for open strings 
are obtained.
Furthermore, open/closed string duality is proved from a judicious integration over the open string fields.
In particular, by restriction to the case of independence on continuous open moduli, the shift formulas of 
\cite{Oog} are reproduced and shown therefore to encode the data of a closed string dual.
\end{quotation}
\vfill
\eject

\end{titlepage}

%\begin{abstract}

%\end{abstract}

\section{Introduction}

A proper target space formulation of open plus closed topological strings
is important for several reasons,
the most compelling in our opinion being a better understanding of open/closed string duality which, once an off shell
formulation of the theory is given, should become manifest.
Actually, this is the main subject of this paper.
Open/closed duality is commonly believed
\cite{OV1}
to be the effect of integrating out open strings
in the complete string field theory, leaving then a purely closed string theory
on a suitably modified background.
This program is very hard to be realized in the full string theory, 
but it becomes tractable
in its truncation to its BPS protected sectors, namely in topological string theories
\cite{Witten,BCOV}.
This issue has been investigated by several authors in a first quantized or {\it on shell}
framework.
Actually, the first examples were discussed in terms of geometric transitions \cite{GV}  
which have been extended to the brane sector in \cite{OV1}.
Then, this picture has been refined in terms of a proper world-sheet analysis in \cite{OV2}.
More advances on-shell computations has been prompt by \cite{remodeling} and then further
by \cite{tso} and \cite{emanuel}.
A distinctive feature of topological strings is that
the non-holomorphic dependence of its amplitudes can be
recursively computed by means of the holomorphic anomaly equations (HAE) \cite{BCOV}. 
It turns out that the target space formulation of the closed string in terms of the Kodaira-Spencer gravity
is very effective in reproducing these recurrence relations from a Feynman diagram's expansion.
This also provides a target space interpretation of the various coefficients appearing in the HAE.
These latter have been more recently extended to open strings in \cite{Wal1} and \cite{BT}.
These were further studied in \cite{Alim}. 
The topological open string target space formulation has been actually obtained long ago in \cite{W1}
where it was shown to be given by the Chern-Simons theory for the A-model and its holomorphic version
for the B-model. These are formulated for a fixed on shell background geometry, in particular for the B-model
the holomorphic Chern-Simons is formulated with respect to an integrable complex structure on the Calabi-Yau
target.
Since the aim of this paper is to study a string field theory formulation of topological open plus closed
strings on equal footing, we will extend this framework to non-integrable structures.
The formulation of holomorphic anomaly equations and the
target space interpretation of its structure functions are
very important tools to obtain a well defined computational framework
for open topological strings. D-branes sources for closed strings are actually
represented in the HAE by the Walcher's term \cite{Wal1} whose target space interpretation
has been given in terms of the Griffith's normal function (see also \cite{MW}).
For the B-model this boils down to the on shell holomorphic Chern-Simons action.
A remarkable observation \cite{Oog} consists in the proof that the Walcher's term can be reabsorbed
by a shift in the string coupling constant and the closed moduli. This indeed realizes
an on shell proof of the open/closed duality, although at frozen open moduli.

In the following we will study this problem from a second quantized point of view, which turns out to be
the most appropriate to study open/closed duality in particular for the B-model.
We will work out the BV formulation of the holomorphic
Chern-Simons theory by leaving the gravitational background (Kodaira-Spencer gravity field) off shell.
This allows us to reformulate open-closed duality as a process of partial functional integration
over the open string fields. From the BV viewpoint this procedure
follows by partial integration of a proper
subset of fields and anti-fields of a solution of the BV master equation by which one gets another solution
depending on a reduced set of fields. This is known as Losev trick \cite{Losev}.
In particular, at frozen open string moduli, we will show that this partial integration
exactly reproduces the shift formulas proposed in \cite{Oog}\cite{Wal2}.
More in general, our BV formulation proves the existence of definite shift formulas
also in presence of open moduli providing a computational set-up to determine them.
Moreover, it yields a target space interpretation of the coefficients of the
extended HAE for open string moduli as in \cite{BT}\cite{Wal3}.

The paper is organized as follows. 
In section 2 we discuss the classical complete string field theory action for open plus closed B-model.
In section 3 we proceed to its quantization using the BV formalism.
In section 4 we discuss the target space interpretation of the coefficients in the open HAE from the string field theory.
In section 5 we formulate and prove in general the open/closed duality or the B-model
and apply it to the setting of \cite{Wal1,Oog}.
In section 6 we collect few concluding comments.

\section{Open-closed effective field theory}

It is well known from \cite{BCOV} that the effective space-time theory corresponding to the B-model for closed strings is given by the Kodaira-Spencer 
theory of gravity:
\begin{equation}\label{d}
\lambda^{2}S_{KS} = \int_{X}\frac{1}{2}A'\frac{1}{\partial}\overline{\partial}A' - \frac{1}{3}[(A + x)(A + x)]'(A + x)'
\end{equation}
where $\lambda$ is the string coupling,
$A$ and $x$ are $(0,1)$ forms with values in the $(1,0)$ vector field that is, in coordinates, 
$A = A_{\overline{i}}^{j}dz^{\overline{i}}\frac{\partial}{\partial z^{j}}$ and similarly for $x$. 
In (\ref{d})
$A'=i_A\Omega_0=3(\Omega_0)_{ijk}A^i_{\bar i}dz^j dz^k dz^{\bar i}$ and similarly for $x'$ where
$\Omega_{0}$ is
the holomorphic three form on the Calabi-Yau target space $X$\footnote{Factors may change depending on the conventions; 
we will use the ones of \cite{Tod} and \cite{KNS}.}.
$A + x$ is defined to be a deformation of the complex structure of $X$ split into an infinitesimal part, $x$, and a finite one, $A$. 
The full deformation, $A + x$, is parametrized by the shift 
$\overline{\partial}_{\overline{i}} \rightarrow \overline{\partial}_{\overline{i}} - (x_{\overline{i}}^{j} + A_{\overline{i}}^{j})\partial_{j} $. 
By definition the coefficients of forms with barred indices transform in the same way : 
$w_{\overline{i}} \rightarrow w_{\overline{i}} - (x_{\overline{i}}^{j} + A_{\overline{i}}^{j})w_{j} $. 
In addition  $dz^{j} \rightarrow dz^{j} + ( x^{j}_{\overline{i}} + A^{j}_{\overline{i}} )dz^{\overline{i}}$, while
 $\partial$ and $d\overline{z}$ are fixed (their shift refers to the antitopological theory). 
In this way real objects as the de Rham differential $d$ or a real form $w_{i}dz^{i} + w_{\overline{i}}dz^{\overline{i}}$ remains unchanged. 
The condition of integrability of the modified complex structure is   
\[
0 = (\overline{\partial} - x - A)(\overline{\partial} - x - A) = - \overline{\partial}(A + x) + \frac{1}{2}[A + x,A + x] = 0
\]
which can be rewritten, due to the fact that $\overline{\partial}x = 0$, $x$ being the background parameter valued in 
$H^{0,1}_{\overline{\partial}}(TM)$, as 
\be 
\overline{\partial}A'  = \partial((A + x)\wedge ( A + x ))' .
\label{kseom}\ee
(\ref{kseom}) is the equation of motion of (\ref{d}). 
Let us stress that it is {\it crucial} the fact that $x$ does not appear in the kinetic term of (\ref{d}).
%Howether $x$ does not appear in the kinetic part so it is taken as the background corresponding to an infinitesimal shift and always 
%satisfing the linearized equation $\overline{\partial}x = 0$ ( in fact $x \in H^{0,1}_{\overline{\partial}}(TM)$ to fix the ambiguities 
%in the solution from the symmetries of the theory ). 
In addition $A$ is required to satisfy the so called Tian's gauge, $\partial A' = 0$, 
in order to have a well defined kinetic term. 

The symmetries of (\ref{d}) are the $\Omega_0$ preserving reparameterizations of the complex coordinates
%As a theory of gravity its symmetries are among the diffeomorphism. In particular, with parameter $\chi$, the ones such that 
$z^{i} \rightarrow z^{i} + \chi^{i}(z,\overline{z})$ and $z^{\overline{i}} \rightarrow z^{\overline{i}}$ 
%($\chi^{\overline{i}}= 0$) 
while the condition of being $\Omega_{0}$ preserving reads $\partial\chi' = 0$. 
According to $\overline{\partial} \rightarrow \overline{\partial}  - (A + x)$ and owing to the fact that $x$ is a background, 
$A$ transforms
% under diffeomorphism absorbing all the shifts of the members on the right hand side of the equation that is:
as
\begin{equation}\label{e}
\delta A =  -\overline{\partial}\chi - {\cal L}_{\chi}(A + x) = -\overline{\partial}\chi - [\chi,(A + x)]
\end{equation}
Reinterpreting $\chi$ as a ghost field, 
this transformation can be promoted to a nilpotent BRST if
% squares to zero under \footnote{now $\chi$ is the corresponding ghost }
\begin{equation}
\delta \chi = -\frac{1}{2}{\cal L}_{\chi}\chi = -\chi^{i}\partial_{i}\chi.
\end{equation}

The open effective theory has been analysed by Witten in \cite{W1} and for the B-model it is 
given by the holomorphic Chern-Simons action
\begin{equation}\label{a}
\lambda S_{HCS} = \int_{X}\Omega_{0} Tr(\frac{1}{2}B^{0,1}\overline{\partial}B^{0,1} + \frac{1}{3}B^{0,1}B^{0,1}B^{0,1})
\end{equation}
with $B^{0,1}$ a Lie algebra valued $(0,1)$-form. 

The precise definition of the model has been presented in \cite{Tho}.
Indeed (\ref{a}) is globally ill defined. 
From the Chern-Weil theorem we know that only the difference of two invariant polynomials with respect to two different connections 
$\hat{B}$ and $B_{0}$ (dropping for the moment the label $(0,1)$) is an exact form. So using the reference connection $B_{0}$ we can write
\begin{eqnarray}\label{m}
-\int_{K_{4}}\frac{\Theta}{2} Tr(\hat{F}^{2} - F_{0}^{2}) &=& -\int_{K_{4}}\Theta Tr \;\overline{\partial}(\frac{1}{2}\hat{B}\overline{\partial}\hat{B} + \frac{1}{3}\hat{B}^{3} - \frac{1}{2}B_{0}\overline{\partial}B_{0} - \frac{1}{3}B_{0}^{3}) = \nonumber \\
&=& \int_{X}\Omega_{0} Tr(\frac{1}{2}\hat{B}\overline{\partial}\hat{B} + \frac{1}{3}\hat{B}^{3} - \frac{1}{2}B_{0}\overline{\partial}B_{0} - \frac{1}{3}B_{0}^{3})
\end{eqnarray}
where $K_{4}$ is a fourfold containing $X$ as a divisor while $\Theta$ is a connection of the associated line bundle $\call_X$ so that
$\overline{\partial}\Theta = \Omega_{0}\delta(X)$. 
We expand $\hat B$ with respect to the reference connection as
\[
\hat{B} = B + B_{0} 
\]
so that
(\ref{m}) provides the globally well defined action  
\begin{equation}\label{b}
\lambda S_{HCS} = \int_{X}\Omega_{0} Tr(\frac{1}{2}B^{0,1}\overline{\partial}_{B_{0}^{0,1}}B^{0,1} + \frac{1}{3}(B^{0,1})^{3} + F^{0,2}_{0}B^{0,1})
\end{equation}

with $\overline{\partial}_{B_{0}^{0,1}}\varphi \equiv \overline{\partial}\varphi + [B_{0}^{0,1},\varphi]_{\pm}$ with $\pm$ 
depending on the grade of the form $\varphi$. $B_{0}$ is the open string background for the theory and as such
it obeys the holomorphicity condition $F_{0}^{0,2} = 0$. 
%This would delete the term in $F_{0}$ but for now we will leave it there as it will simplify some formulas later.
The symmetries of (\ref{b}) -- at fixed background $B_0$ -- are given by
%The symmetries are clear from the original expression (\ref{m}) that is
%
%\begin{eqnarray}
%\delta \hat{B}^{0,1} = \overline{\partial}\epsilon + [\hat{B}^{0,1},\epsilon] \nonumber \\
%\delta B_{0}^{0,1} = \overline{\partial}\epsilon_{0} + [B^{0,1},\epsilon_{0}] 
%\end{eqnarray}
%
%Then we fix $\epsilon_{0} = 0$ because we want to interpret $B_{0}$ as a background so
\begin{equation}\label{gt}
 \delta B^{0,1} = \overline{\partial}_{B_{0}^{0,1}}\epsilon + [B^{0,1},\epsilon].
\end{equation}

Now we want to explicitly couple the open theory to the closed field that is we want to deform the complex structure of $X$, 
over which the theory is defined, using the fields $A$ and $x$. 
Of course the closed fields are in general not on shell so the new complex structure (better call it almost complex structure) 
is generically not integrable. 
In addition we want to write the new action with respect to the undeformed complex structure in order to keep the closed field explicit. 
Actually, under the deformation
%This means that the integral will be over the original differentials, $dz^{i}$ and $dz^{\overline{i}}$ and, all the forms in the new complex structure, 
%will be written using the rules already stated. 
$\Omega_{0}$ is mapped to \cite{Tod}
\begin{equation}\label{x}
\Omega = \Omega_{0} + (A + x)' - [(A + x)(A + x)]' - [(A + x)(A + x)(A + x)]'
\end{equation}
which is a $(\tilde{3},\tilde{0})$ form with respect to the new complex structure (from now on always indicated with a tilde) 
while with respect to the old one
decomposes in forms of total degree 3, namely $(p,q)$ forms with $p+q=3$.
We can now deform also the remaining $(0,3)$ part of the action, $L_{CS}^{0,3}$, with 
$ L_{CS}^{0,3} \equiv Tr(\frac{1}{2}B^{0,1} \overline{\partial}_{B_{0}^{0,1}}B^{0,1} + \frac{1}{3}(B^{0,1})^{3} + F^{0,2}_{0}B^{0,1}) $,  
into a $(\tilde{0},\tilde{3})$ form. 
%The result, indeed correct, is not in the form most usefull for our purposes. The point is that the integral over the original differentials selects only $\Omega_{0}$ from $\Omega$ leaving all the closed field dependence in $L_{CS}^{\tilde{0},\tilde{3}}$. Instead let us start with rewriting it as a real form, that is 
In order to keep into account the deformation of the complex structure of the full action the simplest way is to use a real form for 
the Chern-Simons term, rewriting
\be
\int_{X}\Omega^{\tilde{3},\tilde{0}} L_{HCS}^{\tilde{0},\tilde{3}} = \int_{X}\Omega^{\tilde{3},\tilde{0}}
L_{CS}=\int_{X}\Omega^{\tilde{3},\tilde{0}}Tr\left(\frac{1}{2}Bd_{B_{0}}B + \frac{1}{3}B^{3} + F_{0}B\right)
\label{real}\ee
where $B$ is a real Lie algebra valued 1-form on $X$. Indeed,
being $\Omega$ a $(\tilde{3},\tilde{0})$ form, the added piece is zero.
% ( even if integrated with respect to the original differentials as can be explicitly check going in components ). 
However, from the path integral quantization viewpoint, we have to define a suitable measure for the new field component 
$B^{\tilde{1},\tilde{0}}$. We will discuss this issue in the next section by using the Batalin-Vilkovisky formalism.
%Of course as an integral over $M$ the identity is valid but, inside a path integral, we are adding all the zero modes from the integration of the 
%new field $B^{\tilde{1},\tilde{0}}$. 
%We will soon take care of these in a naive way; a complete BV formulation of the open theory 
For the Kodaira-Spencer gravity in antifield formalism see \cite{BCOV}.
% ) and a correct gauge fixing will be done in the next section.
Let us notice that the real form $L_{CS}$ is completely independent from the closed field, while it is $\Omega$ which really takes care to project the 
action onto the new complex structure selecting the complementary form degree from $L_{CS}$. 

%To fix the zero modes we force the $B^{\tilde{1},\tilde{0}}$ field to be a flat connection, that is we add a Lagrange multiplier times the field strenght for $B^{\tilde{1},\tilde{0}}$, ${\bar\psi} F_{B_{0}}^{\tilde{2},\tilde{0}}$, with $F_{B_{0}}\equiv d_{B_{0}}B + B^{2} + F_{0}$\footnote{ being $F_{0}^{\tilde{0},\tilde{2}} = 0$ the condition over the background in the new complex structure we could avoid writing $F_{0}$ all over, but its presence is usefull as it mades explicit the transformation of $F_{B_{0}}$ in the adjoint}. Obviously ${\bar\psi}$ is a $(\tilde{1},\tilde{3})$ form. The path integral is now well defined dividing by the ( finite )  number of flat holomorphic connections over $M$. So the action has become 
%\begin{equation}
%\lambda S^{c}_{HCS} = \int_{M}\Omega Tr(\frac{1}{2}Bd_{B_{0}}B + \frac{1}{3}B^{3} + F_{0}B) + {\bar\psi}F_{B_{0}}^{\tilde{2},\tilde{0}}
%\end{equation}

Let us consider the symmetries of (\ref{real}). As far as diffeomorphisms (\ref{e}) are concerned, $\Omega$ in (\ref{x}) transforms as ${\cal L}_{\chi}\Omega$
so that the whole action is invariant under the standard action on $B$, namely $\delta B=-\call_\chi B$.

The situation for the Chan-Paton gauge symmetry is more subtle. Indeed,
%The closed symmetries (\ref{e}) are symmetries also for this action, and this is easily seen using the quite surprising result that under the transformations (\ref{e}) $\Omega$ transforms exactly as ${\cal L}_{\chi}\Omega$ \footnote{It can be shown writing ${\cal L}_{\chi}\Omega = d(i_{\chi\Omega}) + i_{\chi}(d\Omega) = (\partial + \overline{\partial})(i_{\chi\Omega}) + i_{\chi}((\partial + \overline{\partial})\Omega)$ and comparing it to the direct transformation of the $A$ field in $\Omega$. Use of the Tian's gauge $\partial A' = \partial\chi' = 0$ should be done} ( clearly all the other fields will transform as usual Lie derivatives and ${\bar\psi}$ transforms such that ${\bar\psi}F_{B_{0}}^{\tilde{2},\tilde{0}}$ is a total Lie derivative. In addition the number of holomorphic connections over $M$ is independent by the closed field because $A$ and $x$ factors out from the condition $F_{B_{0}}^{\tilde{2},\tilde{0}} = 0$ ). 
being the field $A$ off shell, we do not have $d\Omega = 0$. In fact it can be shown \cite{Tod} that $d\Omega = 0$ is equivalent to the equations of motion 
for the Kodaira-Spencer action, $\overline{\partial}A' = \partial((A + x ) \wedge ( A + x ))'$. 
So we expect a variation of the action under the gauge transformations (\ref{gt}) proportional to it. 
We find 
%\footnote{note that the condition $F_{B_{0}}^{\tilde{2},\tilde{0}} = 0$ is stable also under the gauge transformations as $F_{B_{0}}$ transforms in the adjoint. }
\begin{equation}
\delta S_{HCS} = \frac{1}{\lambda }\int_{X}\Omega Tr\; d(\frac{1}{2}\epsilon d_{B_{0}}B + F_{0}\epsilon)
\end{equation}
We can save the day by adding to the action the term $-\frac{1}{2}\Omega db$, where $b$ is a real 2-form field transforming as \cite{BW}:  
\begin{eqnarray}
\delta B &=& d_{B_{0}}\epsilon + [B,\epsilon] \nonumber \\
\delta b &=& Tr(\epsilon d_{B_{0}}B +2 F_{0}\epsilon) 
\end{eqnarray}
%
%This turns out to be the same $b$ field of \cite{BW} and \cite{DGNV}.
The field $b$ acts as a Lagrange multiplier enforcing the Kodaira-Spencer equations for the closed field $A$. 
%So we expect that a path integral over $A$ should be reduced to the single contribution, $A_{os}$, given by the solution of the Kodaira-Spencer equations; 
%said in another way it reduces to the deformation of the original complex structure 
%( $t,\overline{t} \rightarrow t + x + A_{os}(x;t,\overline{t}),\overline{t} $ ) which becomes, because of the equations satisfied by $A$, 
%a new ( integrable ) complex structure. 
However the role of implementation of the associated delta function 
%played by the $b$ term in the action 
requires also a determinant factor such that
\begin{equation}\label{fp}
\int {\cal D}A{\cal D}b e^{-\frac{1}{2}\int_{X}\Omega db} det_{FP}  = 1
\end{equation}
This determinant measure has to be included in the very definition of the theory and will be explicitly derived in the next section.

This isn't really the end of the story as $b$ has shift symmetries along its $(\tilde{2},\tilde{0})$ and $(\tilde{1},\tilde{1})$ components.
%two zero modes to fix: $b^{\tilde{2},\tilde{0}}$ and $b^{\tilde{1},\tilde{1}}$. 
In addition we should specify the full nilpotent symmetries and the gauge fixing. This will be the subject of the next section. 

Summarizing, the classical action for open and closed B-model is
\begin{equation}\label{f}
S_{tot}= \frac{1}{\lambda^{2}}\int_{X}\left(\frac{1}{2}A'\frac{1}{\partial}\overline{\partial}A' - \frac{1}{3}[(A + x)(A + x)]'(A + x)' \right)+
\end{equation}
\[
+ \frac{1}{\lambda}\int_{X} \Omega Tr(\frac{1}{2}Bd_{B_{0}}B + \frac{1}{3}B^{3} + F_{0}B) -\frac{1}{2}\Omega db 
\]

\section{On the BV quantization of Holomorphic Chern-Simons}

In this section we provide the BV action for the holomorphic Chern-Simons theory and a non singular gauge fixing fermion. 
For simplicity in this section we will drop the tilde in the notation for forms in the new complex structure. 
Still the coupling with the closed field is always present.

The classical action is
\be
\lambda S_o=\int_X \Omega^{(3,0)} \left[Tr\left(\frac{1}{2} B d_{B_0} B + \frac{1}{3} B^3 + B F_0\right)-\frac{1}{2}db\right]
\label{hcsa}\ee
This is invariant under the infinitesimal gauge transformations
\bea
s B &=& d_{B_0}\epsilon +[B,\epsilon]+\psi^{(1,0)} \nonumber\\
s b &=& Tr\left(Bd_{B_0}\epsilon+2F_0\epsilon\right) + d\gamma +\eta^{(1,1)}+\eta^{(2,0)}
\label{gsym}\eea
where $\epsilon$ is the usual gauge symmetry ghost while $\psi^{(1,0)}$, $\eta^{(2,0)}$ and $\eta^{(1,1)}$
are the ghosts for the shift symmetries.

By further defining
\bea
s\epsilon &=& -\epsilon^2 \nonumber\\
s\psi^{(1,0)} &=& \left[\epsilon,\psi^{(1,0)}\right] \nonumber\\
s\gamma^{(1,0)} &=& n^{(1,0)}- Tr\left(\epsilon \partial^{(1,0)}_{B_0}\epsilon\right)\nonumber\\
s\gamma^{(0,1)} &=& \partial^{(0,1)}m - Tr\left(\epsilon \partial^{(0,1)}_{B_0}\epsilon\right)\nonumber\\
s\eta^{(1,1)} &=& -Tr\left(\psi^{(1,0)}\partial^{(0,1)}_{B_0}\epsilon\right) -\partial^{(1,0)}\partial^{(0,1)} m
-\partial^{(0,1)} n^{(1,0)}\nonumber\\
s\eta^{(2,0)} &=& -Tr\left(\psi^{(1,0)}\partial^{(1,0)}_{B_0}\epsilon\right) -\partial^{(1,0)} n^{(1,0)}\nonumber\\
sn^{(1,0)}&=&0\nonumber\\
sm &=& 0
\label{pseudobrs}\eea
we get a {\it pseudo}-BRST operator.
Actually the operator $s$ defined by (\ref{gsym}) and (\ref{pseudobrs}) is nilpotent only on shell.
Explicitly, one gets
\be
s^2 b^{(0,2)}=\left(\partial^{(0,1)}\right)^2 m
\label{nnp}\ee
which is vanishing only on shell w.r.t. $b$. 
Actually, as discussed in \cite{Tod}, the differential of the shifted 3-form (\ref{x}) is proportional to the Nijenhuis tensor.
Thus (\ref{nnp}) is proportional to the equation of motion of $b$.
On all other fields one gets $s^2=0$.

The BV recipe is in this case still simple, since one can check that second order in the antifields 
already closes in this case.
By labeling all the fields entering 
(\ref{gsym}) and (\ref{pseudobrs}) as $\phi^i$, we have therefore
\footnote{Here we use the $\lor$-operator as in \cite{BCOV} so that the $\lor$ of a $(3,p)$-form is a $(0,p)$-form.}
\be
S_{BV}= S_o + \int_X \sum_i \phi^*_i s\phi^i + c \int_X \left((b^*)^{(2,2)}\partial^{(1,0)}m\right)^{\lor} (b^*)^{(3,1)}
\label{BV}\ee
where $c$ is a non zero numerical constant which will not be relevant for our calculations (see later).
One can explicitly show that $S_{BV}$ satisfies $\Delta S_{BV}=0$, where $\Delta$ is the BV-laplacian
and $(S_{BV},S_{BV})=0$ the corresponding bracket.
In our conventions, all antifields have complementary form degree with respect to fields.

Let us notice that a parallel result has been obtained in \cite{camillo} by C.~Imbimbo for the A-model. 
Indeed, also in the case of the real Chern-Simons theory, the coupling with the gravitational background requires the 
use of the full BV formalism giving rise to quadratic terms in the anti-fields.

While gauge fixing, we need to add the anti-ghost multiplets for all gauge fixed parameters.
Actually we are going to gauge fix our theory only partially, that is we will keep the 
($\epsilon$-)gauge freedom relative to the Chan-Paton bundle.
By introducing the relevant anti-ghost multiplets, we define the gauge fixing fermion
\bea
\Psi=
\int_X \left\{
{\bar\psi}^{(1,3)}\left(d_{B_0}B + B^2 + F_0\right)^{(2,0)}
+{\bar\eta}^{(2,2)}b^{(1,1)}
+{\bar\eta}^{(1,3)}b^{(2,0)}
\right.\\\left.
+{\bar n}^{(2,3)}\gamma^{(1,0)}
+{\bar m}^{(3,3)} \left(\partial^{(0,1)}\right)^\dagger \gamma^{(0,1)}
+{\bar\gamma}^{(3,2)}\left[\left(\partial^{(0,1)}\right)^\dagger b^{(0,2)}+\partial^{(0,1)}p\right]
\right\}
\eea
by adding the anti-ghost (trivial) part of the BV action in the usual form.
We extend therefore the s-operator action, that is the BV-bracket with the part of the BV action linear in the anti-fields, 
to the anti-ghosts in the trivial way, namely for any anti-ghost $\bar\psi$ we have
$s\bar\psi=\Lambda_{\bar\psi}$ and $s\Lambda_{\bar\psi}=0$. The anti-ghost gauge freedom is fixed by the addition of the 
relevant further sector.

Finally we can compute the (partially) gauge fixed action by specifying all anti-fields as derivatives with respect to 
their relative fields of gauge fermion $\Psi$.
All in all, the (partially) gauge fixed action reads
\bea
 S_{g.f.}=  S_o + s\Psi + c\int_X\left(
{\bar\eta}^{(2,2)}\partial^{(1,0)}m\right)^{\lor}\left(\partial^{(0,1)}\right)^\dagger{\bar\gamma}^{(3,2)}
\label{gfa}
\eea

Let us now perform the path-integral in the different sectors (by naming them by the relative anti-ghost as appearing in 
the gauge fermion).
\begin{itemize}
\item The ${\bar\psi}^{(1,3)}$ is seen to decouple since 
$$s\left\{d_{B_0}B + B^2 + F_0\right\}^{(2,0)}=\partial^{(1,0)}_{B_0}\psi^{(1,0)}+\left[B^{(1,0)},\psi^{(1,0)}\right]_+$$
Therefore we get the contribution 
$$
\int \cald[B^{(1,0)}]\delta\left(\partial^{(1,0)}_{B_0}B^{(1,0)} + B^{(1,0)}B^{(1,0)} + F_0^{(2,0)}\right)
{\rm det'}\left\{\partial^{(1,0)}_{B_0}+\left[B^{(1,0)},\cdot\right]_+\right\}
$$
which counts the volume of the space of holomorphic connections.

\item The two $\bar\eta$-sectors are just algebraic and give a constant contribution to the path-integral.
Notice that while integrating over $\bar\eta^{(2,2)}$ also the last term in (\ref{gfa}) gets involved being reabsorbed in a shift of $\eta^{(1,1)}$.
This gauge fixing of course restricts the field $b$ to be a $(0,2)$-form only and set to zero $\eta^{(1,1)}$ and $\eta^{(2,0)}$.

\item The ${\bar n}^{(2,3)}$ sector is algebraic too and simply sets to zero $\gamma^{(1,0)}$ and its partner. 
%It contributes a multiplicative $1$ to the path integral.

\item The last part is the standard term for higher form BV quantization (see for example \cite{HT}).
The fermionic bilinear operator reduces to 
$$
\calb=
\left(\begin{matrix}-{\partial^{(0,1)}}^\dagger\partial^{(0,1)} & -\partial^{(0,1)} \\
                {\partial^{(0,1)}}^\dagger & 0 \end{matrix}\right)$$ 
mapping $\Omega^{(0,1)}(X)\oplus \Omega^{(0,0)}(X)$ to itself.
The bosonic bilinear operator is instead the anti-holomorphic laplacian 
$\Delta^{(0,0)}={\partial^{(0,1)}}^\dagger\partial^{(0,1)}$ on the scalars $\Omega^{(0,0)}(X)$.
One therefore stays with the gauge fixed measure 
\be
\int \cald[Y] e^{-\frac{1}{2}\int_X Y \calc Y +\int_X J^tY}
\label{circa}\ee
where
$Y=\left(p,\Lambda_{\bar\gamma},b^{(0,2)}\right)$, 
$$
\calc=
\left(
\begin{matrix}
0               & -\partial^{(0,1)} & 0           \\
\partial^{(0,1)}&      0            & {\partial^{(0,1)}}^\dagger\\
0               & -{\partial^{(0,1)}}^\dagger & 0
\end{matrix}
\right)
$$
and the source $J=(0,0,d\Omega)$ takes into account the classical action.
Eq.(\ref{circa}) can be integrated being a Gaussian.

\end{itemize}

Therefore, all in all, we find that the quantum measure for the holomorphic Chern-Simons theory
is
\be
\frac{det'[\calb]}{det'[\Delta^{(0,0)}]\left(det'[\calc]\right)^{1/2}}
e^{J^t(\calc)^{-1}J}
\label{qm}\ee
for a (generically non integrable) almost complex structure.
The determinant of the operator $\calc$ is easily obtained by noticing that
$$
\{\calc,\calc^\dagger\}=
\left(
\begin{matrix}
\Delta^{(0,0)}               & 0            & \left(\partial^{(0,1)}\right)^2 + \left({\partial^{(0,1)}}^\dagger\right)^2\\
0                            & 2 \Delta^{(3,2)}                 &                           0           \\
       \left(\partial^{(0,1)}\right)^2 + \left({\partial^{(0,1)}}^\dagger\right)^2                 &0            & \Delta^{(2,0)} 
\end{matrix}
\right)
$$

We want to compare our open theory, defined as coupled to the closed field $A$, with the standard holomorphic Chern-Simons, 
defined for an integrable complex structure. In particular the two theories should match once we put on shell the closed field. 
So the integral of all the additional fields should contribute as one. Notice that, if the complex structure is integrable, 
then $d\Omega=0$ and the source term is not contributing.
On top of it, since $\left(\partial^{(0,1)}\right)^2=0$, the bosonic operator block-diagonalizes.
Moreover, in this case, the determinant of the fermionic operator $\calb$ can be easily computed
\footnote{This can be done by writing the eigenvector equation for $\calb$ as $\calb \tbinom{a}{b}=\lambda\tbinom{a}{b}$
and then expanding the 1-form $a=\partial^{(0,1)}x+{\partial^{(0,1)}}^\dagger y$ in exact and co-exact parts. 
Then one finds that $b=\lambda x$ and that the eigenvalues of $\calb$ coincide with 
those of $\Delta^{(0,2)}$ for $x=0$ or with the square roots of those of $\Delta^{(0,0)}$ for $y=0$.}
to be equal to $det'\Delta^{(0,2)} \left(det'\Delta^{(0,0)}\right)^{1/2}$.

All in all, we find an overall 
\be
\frac{det'[\Delta^{(0,2)}] \left(det'[\Delta^{(0,0)}]\right)^{1/2}}
{det'[\Delta^{(0,0)}]\left\{\left(det'[\Delta^{(0,2)}]\right)^2
det'[\Delta^{(0,0)}]\right\}^{1/2}
}=\frac{1}{det'[\Delta^{(0,0)}]}
\label{ciapa}\ee
This determines the value of the quantum measure introduced in (\ref{fp}).
The factor (\ref{ciapa}) counts the extra degree of freedom introduced by the $b$ field in the theory.
Indeed the three components of $b^{(0,2)}$ are subject to the gauge freedom by the shift 
of an exact $\partial^{(0,1)}\gamma^{(0,1)}$ term up to the ghost-for-ghost shifting
$\gamma^{(0,1)}$ by $\partial^{(0,1)}m$. Therefore the overall counting is $3-3+1=1$
complex modes.

\section{String field theory as generating function of open and closed HAEs.}

Our claim of having found the effective space-time theory for the open B-model should be checked explicitly. 
Because of tadpole cancellation, see \cite{nostro} and \cite{Wal3}, we know that the open theory is completely well defined only 
in its unoriented version ( as in the case of usual string theories ), so the most general case to consider is for open 
( and closed ) unoriented strings. Closed moduli are known to be unobstructed and so expansions of the amplitudes in their 
value is always possible. 
%For open moduli the claim in \cite{Wal1} is to be generically obstructed apart from specific cases. 
%Nethertheless from the field theory side we don't see any specific reason to treat them differently from the closed ones so we 
%will consider also an expansion around their value. If this can not be the case you can accomodate the most restricting assumption 
%of obstructedness just replacing the expansion in their continuos parameters with, at most, a sum other discrete variables. 
We will proceed similarly for open moduli.
An important result of \cite{BCOV} is that the partition function of Kodaira-Spencer theory encodes the recurrence relations of HAE via 
its Feynman diagram expansion.
%for the topological string can be used to reproduce, 
%inside a master equation, the Holomorphic Anomaly Equation ( from now simply H.A.E. ). 
The generating function of the full HAE of \cite{BT} generalized to the unoriented case should be: 
\begin{equation}\label{t}
e^{W(x,u;t,\overline{t})}  \sim \exp\left(\sum_{g,h,c,n,m} 
\frac{\lambda^{2g - 2 + h + c}}{2^{\frac{\chi}{2} + 1}\;n!m!}
{\cal F}^{(g,h,c)}_{i_{1}\dots i_{n}\alpha_{1}\dots \alpha_{m}}x^{i_{1}}\dots x^{i_{n}}u^{\alpha_{1}}\dots u^{\alpha_{m}}\right)
\end{equation}
up to an overall $\lambda$ dependent prefactor which encodes the contact terms in one loop calculations and will be discussed later.
This prefactor $\lambda^{\dots}$  is encoded, in the field theory side, in the measure of the path integral, namely as the 
multiplicative term weighting the regularized determinants with omitted zero modes. 
From now on, in any case, we will focus on the perturbative expansion in $\lambda$.

%For the specific cases which arise at one loop a slight modification appears in the shape of a prefactor in front of the exponential, $\lambda$, erased to some power. This power is defined to get the right value of the contact terms in the one loop H.A.E. We will not need it in the general case, so we leave it unspecified ( and it is its missing which explains the presence of $\sim$ in ( \ref{t} ) ). Howether in a few paragraphs it will appear in a specific situation.

The notation is as follows: ${\cal F}^{(g,h,c)}_{i_{1}\dots i_{n}\alpha_{1}\dots \alpha_{m}}$ is 
the string amplitude with genus $g$, $h$ boundaries, 
$c$ crosscaps, $n$ marginal operator insertions in the bulk and $m$ on the boundary.
% ( the last ones made by operators parametrising shifts in the Wilson lines generating open moduli ). 
%The parameter $\lambda$ is the string coupling constant, 
The $x^{i}$'s are the expansion coefficients of $x$ in a base of Beltrami differentials, $x = x^{i}\mu_{i}$ 
and the $u^{\alpha}$'s are the expansion coefficients for $B_{0}$ in a basis $T_\alpha(x)$ of the open moduli 
$H^{(0,1)}(X,Ad_E)$, namely $B_{0} = u^{\alpha}T_{\alpha}$.
%with $T_{\alpha}$ a base of Lie algebra matrices. 
Thus the fields appearing as backgrounds in the field theory are the open and closed moduli themselves. 
%Note the dependence of $W$ by $t$ and $\overline{t}$ parametrising the original complex structure background shifted ( infinitesimally ) as $t \rightarrow t + x$ and $\overline{t} \rightarrow \overline{t}$ \footnote{these are the coordinates called Canonical in \cite{BCOV} and they have the special property to reduce all the covariant derivatives to ordinary ones; this is the reason why in the following insertion of operators on the Riemann surface is obtained by a simple derivative.}, and the dependence of $B_{0}$ by $x$ due to the condition it should satisfy as a background, $F_{0}^{\tilde{0},\tilde{2}} = 0$, which depends explicitly by $x$. 
The factor $\frac{1}{2^{\frac{\chi}{2} + 1}}$ is explained in \cite{Wal3} and obviously $\chi = 2g -2 + h + c$. 
If what we are doing is consistent it should be true that 
\begin{equation}\label{g}
\int {\cal D}A{\cal D}B{\cal D}b\dots e^{-S_{tot}(x,B_{0}(x);t,\overline{t};A,B,b,\dots)} = e^{W(x,u(x);t,\overline{t})}.
\end{equation}
%where the massive fields has been integrated out leaving only the background field dependence in (\ref{t}). The dots in the path  integral are there to replace the missing fields in the completly gauge fixed action. 
We want to compare this at tree level, that is at $g = 0, h = 0,1, c = 0$ and $g = 0, h = 0, c = 1$, and obtain in this way some explicit 
expressions for all the basic objects entering the extended HAE of \cite{BT} computed at a generic background point. 
These amplitudes are already known and computed by worldsheet methods and the two results should of course match. 
To this end we will differentiate, at each order in $\lambda$, both members with respect to the moduli parameters $x^{i}$ and $u^{\alpha}$
and identify the corresponding coefficients. 

%Three 
%Four 
A comment is in order. 
%First, following \cite{Oog} and \cite{Wal2} it seems we should have added in (\ref{t}) a new independent 't Hooft like parameter 
%$\beta$ in the form $\beta^{h}$ to weight indipendently the number of boundaries. 
%Howether, as explained in \cite{nostro} and \cite{Wal3}, $\beta$ spoils the important fact, due to tadpole cancellation, 
%that only the sum of certain amplitudes, as for example the cylinder, the M$\ddot{o}$bius and the Klein bottle, are well defined and meaningful. We will nevertheless introduce it in (\ref{z}) to follow the setup of \cite{Oog} and \cite{Wal2} not caring for the moment about the issue of tadpole cancellation. In that case to catch the right expansion in $\beta$ we will formally multiply the Chern-Simons action by $\beta$ itself; working with frozen open moduli the trace will be sobstituted by a factor of $N$ ( Chan-Paton factors ) multipling the open field expression ( without Lie algebra matrices ); so $\beta$ will simply be interpreted as $N$ itself.
%Second, note that the expressions obtained from the right hand side of (\ref{g}) are explicitly the Taylor expansions around $x^{i} = 0$, $u^{\alpha} = 0$ of the amplitudes without the $x^{i}$ and $u^{\alpha}$ insertions, as it is clear given (\ref{t}), while, usually, the quantities we are going to obtain are computed in $x = u = 0
We should remember that the expression (\ref{t}) is the partition function for the unoriented theory. 
As explained in \cite{Wal3} this differs from the oriented one simply projecting the space of operators in the theory 
to the unoriented sector that is the ones with eigenvalue $+1$ under the parity operator $\cal{P}$. 
Being these operators nothing else than deformations of the moduli space of the theory,  
we have to consider only its invariant part under $\cal{P}$ and then parametrise with $x^{i}$ and $u^{\alpha}$ its tangent space. 
This means that the $x^{i}$'s and the $u^{\alpha}$'s appearing in ( \ref{t} ) are really a subset of the ones in the oriented case. 
Specifically it implies a restriction on the space of complex structures for what matters $x$ and a reduction to $Sp(N)/SO(N)$ groups for $u$. 
Still some amplitudes, as the sphere with three insertions, are perfectly meaningful also in the oriented case. 
This is why we will generically not specify to which space the $x^{i}$'s and the $u^{\alpha}$'s belongs: 
it is possible to restrict their value depending on the case. 

\subsection{$g = 0$, $h = c = 0$}

Here we start the comparison between the string theory partition function and the space-time path integral (\ref{g}).
We begin from the coefficients at lowest order in $\lambda$. 
From the point of view of (\ref{t}) this is the amplitude at $g = h = c = 0$ with weight $\frac{1}{\lambda^{2}}$; 
on the field-theory side the contribution should come only from the Kodaira-Spencer action, also at weight $\frac{1}{\lambda^{2}}$. 
We know that the right-hand side of equation (\ref{g}) at this order in $\lambda$ has no dependence on open moduli 
(because without boundaries, $h = 0$, there is no space for open operator insertions) and the building block amplitude being $C_{ijk}(x)$:
\[
C_{ijk}(x) = {\cal F}^{(0,0,0)}_{ijk}(x) = \sum_{n} \frac{1}{n!}{\cal F}^{(0,0,0)}_{ijki_{1}\dots i_{n}}x^{i_{1}}\dots x^{i_{n}} 
= \frac{\partial}{\partial x^{i}}\frac{\partial}{\partial x^{j}}\frac{\partial}{\partial x^{k}}W\mid_{(order \lambda^{-2})} 
\]
Being at tree level and given (\ref{g}), the same result can be obtained ( see \cite{BCOV} ) deriving the Kodaira-Spencer action on shell ( $A = A(x)$ ) 
with respect to three $x^{i}$. The three derivative term gives\footnote{The factor $-2$ depends on our 
conventions which are slightly different from \cite{BCOV}.}
\[
-2\int_{M}\left[\left(\mu_{i} + \frac{\partial A(x)}{\partial x^{i}}\right)\wedge \left(\mu_{j} 
+ \frac{\partial A(x)}{\partial x^{j}}\right)\right]' \left(\mu_{k} + \frac{\partial A(x)}{\partial x^{k}}\right)' = C_{ijk}(x)
\]
The only point of possible confusion for the BCOV educated reader both here and in the subsequent computations, comes from the novel cross 
dependence of open and closed field on shell by each other by means of the field equations which are now modified with respect to the ones 
obtained with the open and closed actions separated. This might seem to carry on additional induced derivatives and contributions as, 
in this case, an induced open moduli dependence carried by the on shell closed field which would lead to the paradox of a non vanishing 
amplitude corresponding to a sphere with boundary insertions! 
Fortunately, integrating out the field $b$ does the job of enforcing the closed field solutions that would be obtained from the 
Kodaira-Spencer action alone! 
%This is clear being $d\Omega = 0$ exactly equivalent to $\overline{\partial}A' = \partial(A + x \wedge A + x)'$. 
It will be true instead that the on shell open fields will carry some closed field dependence as the $B$-field equation is: 
$F_{B_{0}}^{\tilde{0},\tilde{2}} \equiv (d_{B_{0}}B + B^{2} + F_{0})\mid^{\tilde{0},\tilde{2}} = 0$ which is both $B_{0}(u)$ and $x$ dependent.

This is a good place to stop and discuss the connections between our result for the coupling between the open theory and the closed one, 
and the comments made by Witten in \cite{W1} about this point. In his paper Witten uses an argument from the fatgraph description of 
a string tree level amplitude to infer that, if one considers a diagram with $n$ bulk and $m$ boundary insertions, 
in general the bulk operators will reduce to exact (with respect to the topological charge) objects and so will decouple. 
This goes through even in the case $m = 0$ as long as some boundaries are present. 
The direct consequence is that the on-shell couplings between closed and open strings are zero.
How can then one justify the non vanishing of the 
$\Delta_{ij}$ amplitudes of \cite{Wal1}, \cite{Wal2} and \cite{Wal3}?
Our answer is in a sense a weakened realization of Witten's idea, still allowing non zero amplitudes with bulk operators and boundaries. 
The key role is played by the field $b$, generated in the action to maintain the gauge symmetries in the Chern-Simons term. 
This field, once it is integrated over, fixes the closed field $A$ to be on shell with respect to the original Kodaira-Spencer equations 
and so defining a shift of an integrable complex structure. 
This translates to the fact that the original genuine coupling between open and closed fields in the action reduces to a coupling 
between an open, integrated field and an on-shell closed field. That is it represent a new Chern-Simons expansion around a new shifted 
and fixed complex structure. So the path integration of the closed field $A$ reduces to a single contribution coming from the unique 
deformation of the original complex structure with respect to which the Kodaira-Spencer action is written, 
this contribution being weighted by the corresponding Kodaira-Spencer on shell action.
If closed strings are substantially decoupled by the open theory, what is then their role? This is the next point discussed by Witten 
in \cite{W1} where their crucial role in anomaly cancellation is pointed out.
%and his claim is that closed strings are there to cure anomalies occurring in the ``open'' field theory. 
For example, in the A-model, whose effective theory is the real Chern-Simons, a well known topological anomaly is present. 
It comes from the $\eta$-invariant of \cite{W2}, whose dependence by the metric is compensated by the addition of a gravitational Chern-Simons. 
Then an additional anomaly connected to the framing of the target space is well known. 
In the case of the B-model however, the $\eta$-invariant is simply zero because the spectrum of eigenvalues of the determinant 
whose phase is $\eta$, is symmetric around zero \cite{Tho}. 
Instead we have one loop anomalies corresponding to a dependence by the wrong moduli \cite{new} (K$\ddot{a}$hler moduli in this case) 
which is cured by tadpole cancellation, \cite{Wal3} and \cite{nostro}, involving unoriented contributions in the closed strings sector (Klein bottle).

\subsection{$g = 0$, $h + c = 1$ }

In this subsection we want to compare the world-sheet and the target space perspective at order $1/\lambda$.
From the string theory side the relevant amplitudes of weight $\frac{1}{\lambda}$ ( $g = 0$, $h + c = 1$ ) entering the HAE
were discussed in \cite{BT}. From the field theory perspective all of them should be reproduced by the holomorphic Chern-Simons action.

Let us start with purely closed moduli dependence. This can come either from both the explicit dependence by $x$ in $\Omega$ and 
by the induced dependence in the $A(x)$ and $B(x,u)$ fields on shell, or implicitly through the background $B_{0}(x)$. 
We will find that the dependence w.r.t. closed moduli explicit and in the on shell fields, both closed and open, 
correspond to bulk insertion in the string amplitude, while the dependence w.r.t. closed moduli in the background open field
corresponds to induced boundary insertions\footnote{An additional closed moduli dependence in the worldsheet action would come 
also from the Warner term \cite{War}. For the B-model this additional boundary term, needed to make the action invariant, vanishes 
under the usual boundary conditions \cite{W1} as discussed in \cite{mirror}.}.  

The two operators will be indicated as $\phi_{i}$ and $\psi_{i}$ (so for example $C_{ijk} = \langle \phi_{i}\phi_{j}\phi_{k}\rangle_{0,0,0}$ 
where the subscript denotes the triple $g,h,c$). 

The first amplitude we want to derive is $\Delta_{ij} = \langle \phi_{i}\phi_{j}^{[1]}\rangle_{0,1,0 \; + \; 0,0,1}$ which 
was computed in \cite{Wal1} and \cite{Wal3} as additional building block for the extended HAE.
This is the disk plus the crosscap with two bulk insertions. 
In particular $\phi_i$ is a local insertion while $\phi_j^{[1]}$ is an integrated one being the second step of the descent equation.
So, from (\ref{f})
% ( remember $Tr(\frac{1}{2}Bd_{B_{0}}B + \frac{1}{3}B^{3} + F_{0}B) \equiv L_{CS}$ )
we get
\begin{equation}
\frac{1}{\sqrt{2}}\Delta_{ij}(x)  = \int_{X} d_{i}d_{j}\Omega L_{CS} + \int_{X} d_{i}\Omega d_{j}L_{CS} + \int_{X} d_{j}\Omega d_{i}L_{CS} + 
\int_{X} \Omega d_{i}d_{j}L_{CS} 
\label{paletta}
\end{equation}
where all the fields are on shell; $d_{i}$ is the derivative with respect to the closed modulus $x^{i}$, both explicitly and 
through the dependence induced by $A(x)$ and $B(u,x)$; the factor $\frac{1}{\sqrt{2}}$ comes from the normalization in (\ref{t}). 
Using the field equations for $B$ we obtain the identity
\[
0 = d_{j}\left(\int_{X} \frac{\delta S_{HCS}}{\delta B}\mid_{B=B(u,x)}d_iB(u,x)\right)= 
d_{j}\left(\int_{X} \Omega d_{i}L_{CS} \right) = \int_{X} d_{j}\Omega d_{i}L_{CS} + \int_{X} \Omega d_{i}d_{j}L_{CS} 
\]
that is, the last two terms in (\ref{paletta}) cancel. This is nothing but Griffith's transversality condition for the normal function as stated in
\cite{Wal1}. So we get
\begin{equation}\label{l}
\frac{1}{\sqrt{2}}\Delta_{ij}(x)  = \langle \phi_{i}\phi_{j}\rangle_{0,1,0 \; + \; 0,0,1} = 
\int_{X} d_{i}d_{j}\Omega L_{CS} + \int_{X} d_{i}\Omega d_{j}L_{CS} 
\end{equation}
This differs from the expression derived in \cite{Wal1,Wal3} by the first term. However notice that (\ref{l}) is valid 
at a generic value $x$ for closed string moduli, 
while the ones of \cite{Wal1,Wal3} are evaluated at $x = 0$, where the double derivative of $\Omega$ is vanishing. 
This comes from expression (\ref{x}) and from the fact that 
$A(x) = O(x^{2})$ as follows by solving the Kodaira-Spencer equations iteratively.
  
Let us now consider the amplitudes with one bulk and one boundary insertion. The latter, as already stated, is obtained 
from the derivative with respect to 
the background open field $B_0$ which depends on $x$:
\[
\frac{1}{\sqrt{2}}\Delta'_{ij} = \langle \phi_{i}\psi_{j}^{[1]}\rangle_{0,1,0} = 
\left(d_j B_{0}(x) \frac{\delta}{\delta B_{0}(x)}\right)d_{i}S_{HCS}
\]
To compute this term from the space-time point of view it is easier to start from the action written in terms of 
$\hat{B}$ and $B_{0}$ (\ref{m}). The result follows immediately as 
\begin{equation}
\frac{1}{\sqrt{2}}\Delta'_{ij} = \langle \phi_{i}\psi_{j}^{[1]}\rangle_{0,1,0} = -\int_{X} d_{i}\Omega Tr(d_{j}B_{0}(x)F_{0}) 
\end{equation}
once the e.o.m. of the open field are imposed.
%where the equation of motion for $B$ avoid a second term with $d_{i}$ acting over $L_{CS}$ ( here rewritten in terms of $\hat{B}$ and $B_{0}$ ).

Now we pass to the purely open moduli derivatives. The only term is the one derived three times or, 
equivalently, the one with three boundary operator insertions: 
$\Delta_{\alpha\beta\gamma}$. Again using the form (\ref{m}) we need only explicit derivatives with respect to $u^{\alpha}$ 
(remind that $B_{0} = u^{\alpha}T_{\alpha}$ ). The result is 
\begin{equation}\label{n}
\frac{1}{\sqrt{2}}C_{\alpha\beta\gamma} = \langle \Theta_{\alpha}\Theta_{\beta}\Theta_{\gamma}\rangle_{0,1,0} 
= -\int_{X}\Omega Tr(T_{\alpha}T_{\beta}T_{\gamma}) 
\end{equation}
which is the same that would be derived with worldsheet methods in analogy to $C_{ijk}$.

Finally we have mixed terms. These are similarly obtained giving
\begin{equation}\label{o}
\frac{1}{\sqrt{2}}\Pi_{\alpha i} = \langle \Theta_{\alpha}\phi_{i}\rangle_{0,1,0} = -\int_{X} d_{i}\Omega Tr(T_{\alpha}F_{0}) 
\end{equation}
and
\begin{equation}\label{p}
\frac{1}{\sqrt{2}}\Delta'_{\beta i \alpha } = \langle \Theta_{\beta}\psi_{i}^{[1]}\Theta_{\alpha}\rangle_{0,1,0} = -\int_{X} 
\Omega Tr(T_{\beta}d_{i}B_{0}T_{\alpha}) 
\end{equation}
%The only ingredient missing seems the one containing the broken supersymmetric current $B_{\beta i \alpha}$. Still this always comes added to $\Delta'_{\beta i \alpha }$ and from the point of view of the right hand side of (\ref{g}) it should be contained in the same term as  $\Delta'_{\beta i \alpha }$ so probably it is already inside (\ref{p}).

\section{Open-Closed string duality as a Losev trick}
\label{zio}

Let us explain a basic argument about open-closed string duality in second quantization.
This is referred to the topological string theory at hand (B-model), but in principle should hold 
in a more general setting.

The Losev trick, as explained in \cite{Losev}, consists in a procedure to obtain solutions of the 
quantum Master Equation in Batalin-Vilkovisky quantization by partial gauge fixing.
In its generality it reads as follows.
Let $S(\Phi,\Phi^*)$ be a solution of the quantum Master equation 
\be
\Delta\left(e^{-S/\hbar}\right)=0
\label{qme}\ee
where $\Delta=\partial_\Phi\partial_{\Phi^*}$ is the nilpotent BV laplacian.
Suppose that the fields/anti-fields space $\calF$ is in the form of a 
fibration
$$
\begin{matrix}
\calF_2 & \hookrightarrow & \calF \\
\,      & \,              & \downarrow \\
\,      & \,              & \calF_1
\end{matrix}
$$
so that one can choose a split coordinate system $(\Phi,\Phi^*)=(\Phi_1,\Phi^*_1,
\Phi_2,\Phi^*_2)$ such that the BV laplacian splits consistently as
$\Delta=\Delta_1 + \Delta_2$ with $\Delta_1^2=0$.
Then, assuming the existence of a non singular gauge fermion $\Psi$, 
one can consider the partially gauge fixed BV effective action
\be
e^{-\frac{1}{\hbar} S_{eff}(\Phi_1,\Phi^*_1)}
=
\int_{\calF_2}\cald\left[\Phi_2,\Phi^*_2\right] e^{-\frac{1}{\hbar} S(\Phi,\Phi^*)}
\delta\left(\Phi_2^*-\partial_{\Phi_2}\Psi\right).
\ee
which can be readily seen to satisfy the reduced BV Master equation
\be
\Delta_1 e^{-\frac{1}{\hbar} S_{eff}(\Phi_1,\Phi^*_1)}=0
\label{qme1}\ee
Actually -- the proof is two lines -- one consider (\ref{qme}) partially
gauge fixed on the fibers and integrated along the fiber $\calF_2$
$$
0=\int_{\calF_2}\cald\left[\Phi_2,\Phi^*_2\right] \left\{\Delta_1+\Delta_2\right\}
e^{-\frac{1}{\hbar} S(\Phi,\Phi^*)}
\delta\left(\Phi_2^*-\partial_{\Phi_2}\Psi\right)
= 
\Delta_1 e^{-\frac{1}{\hbar} S_{eff}(\Phi_1,\Phi^*_1)} + $$ $$+
\int_{\calF_2}\cald\left[\Phi_2\right] 
\left\{
\frac{d}{d\Phi_2}
           \left( 
           \left[\partial_{\Phi_2^*}
            e^{-\frac{1}{\hbar} S(\Phi_1,\Phi_2,\Phi^*_1,\Phi_2^*)}
           \right]_{\Phi_2^*=\partial_{\Phi_2}\Psi}
	   \right)
           - 
           \partial^2_{\Phi_2}\Psi \cdot \left(\partial^2_{\Phi_2^*}
e^{-\frac{1}{\hbar} S(\Phi,\Phi^*)}\right)\mid_{\Phi_2^*=\partial_{\Phi_2}\Psi}
\right\}$$
Now, the last line vanishes because of translation invariance of the path-integral along the fiber
and field/anti-field opposite statistics, so that we recover (\ref{qme1}).
Let us notice that the resulting BV effective action depends on the particular gauge fixing 
chosen to integrate the fiber degrees of freedom. This dependence is BV trivial in the effective action.

Let us now specify the above setup to open/closed string theory, namely we identify
$\calF$ with the open and closed string theory, $\calF_2$ with the open strings
and $\calF_1$ with closed strings.
The complete theory is given by the BV action 
\be
S_{c+o}(A,B;x,u,\lambda)=S_c(A;x,\lambda)+S_o(A,B;x,u,\lambda)
\label{c+o}\ee
where 
%\begin{itemize}
%\item $A$ are the fields/anti-fields in the closed string sector,
%\item $B$ are the fields/anti-fields in the open string sector,
%\item $x$ are the closed string background moduli,
%\item $u$ are the open string background moduli,
%\item $\lambda$ is the string coupling constant,
$S_c(A;x,\lambda)$ is the closed string BV action, and 
$S_o(A,B;x,u,\lambda)$ completes the open and closed BV action.
%\end{itemize}
The BV laplacian takes the form $\Delta_{c+o}=\Delta_c + \Delta_o$.
We assume that both closed and open plus closed strings have been BV formulated, so that
the corresponding quantum Master equations hold. 
Moreover, the uniqueness of closed string field theory is taken to mean that all solutions of 
the quantum Master equation, with proper boundary conditions in the string coupling dependence -- namely
the background independence of the kinetic term, 
are given by $S_c(A;x,\lambda)$ for some background $x$ and the choice of the string coupling constant $\lambda$.
For the B-model, this is explicitly proved in \cite{BCOV}.

Therefore, by specifying the Losev trick to our case, we obtain that
the effective action obtained from (\ref{c+o}) by partial gauge fixing and integration over the open string 
field, satisfies the quantum Master equation (\ref{qme1}) that is the quantum master equation for the {\it closed} 
string field theory.
Notice that, by definition,
\be
e^{-S_{eff}(A,x,\lambda,u)}=
e^{-S_c(A;x,\lambda)}
\int_{\tiny
\begin{matrix} gauge \\ fixed \end{matrix}
}
\cald[B]e^{-S_o(A,B;x,u,\lambda)}
\label{yyy}\ee
approaches the required boundary condition in the string coupling constant dependence.
The actions entering (\ref{yyy}) are required to have a canonically normalized kinetic term.
Therefore, we conclude that the effective action (\ref{yyy}) has to be
the closed string field action (in some gauge determined by the gauge fixing in the open string sector)
for a shifted set of background moduli and a redefined string coupling constant, that is,
\be
\caln\,\, e^{-S_{c}(A;x^\star,\lambda^\star)}=
e^{-S_c(A;x,\lambda)}
\int_{\tiny
\begin{matrix} gauge \\ fixed \end{matrix}
}
\cald[B]e^{-S_o(A,B;x,u,\lambda)}
\label{yygen}\ee
up to a field independent normalization $\caln$.

The particular case we have in mind is therefore the topological B-model, where
$S_c$ is the Kodaira-Spencer gravity action 
and $S_o$ the holomorphic Chern-Simons action 
suitably coupled to the Kodaira-Spencer field as discussed in the previous sections.
After passing to flat coordinates, (\ref{yygen}) then specifies to
\be
\caln(u,x,\lambda^{-1}\Omega_0)\,\, e^{-\frac{1}{{\lambda^\star}^2}S_{KS}(A^*,x^\star)}=
e^{-\frac{1}{\lambda^2}S_{KS}(A,x)}
\int_{\tiny
\begin{matrix} gauge \\ fixed \end{matrix}
}
\cald[B]e^{-\frac{1}{\lambda}S_{HCS}(A,B,x,u)}
\label{yy}\ee
where 
the closed string field gets renormalized as $A^\star/\lambda^\star=A/\lambda$.
In (\ref{yy}) $\caln$ is a normalization factor\footnote{The particular dependence on the ratio $\Omega_0/\lambda$ is due to the fact that 
we have chosen flat coordinates $u,x$ for the moduli. See next section for a specific discussion on the relevance of the normalization factor 
in comparing with \cite{Wal2}.} and
\be
\frac{1}{\lambda^\star}=\frac{1}{\lambda} + \delta(u,x,\lambda)
\quad {\rm and} \quad
(x^\star)^i=x^i +\delta^i(u,x,\lambda)
\label{shift}\ee
are some shifted background and string coupling. All these are {\it to be determined} 
and can be perturbatively computed from (\ref{yy})
by Feynman diagrams expansion or with non perturbative techniques when available.
The redefinition (\ref{shift}) is a generalization (with tunable open moduli) of the moduli shift in \cite{Oog}.
The aim of the next subsection is to show that, at frozen open moduli, the above formulas reproduce the shift of \cite{Oog}.

\subsection{Open-closed duality at frozen open moduli}\label{z}

In this subsection we want to apply the general arguments just explained in Section \ref{zio} to the oriented string theory
with frozen open moduli \cite{Oog}. Indeed, since we will work just at tree level, we do not have to deal with unoriented amplitudes.
The effect of freezing the open moduli is easily obtained by replacing the non abelian field $B$ with $N$ identical copies of an abelian one, 
reducing the trace simply to a Chan-Paton factor $\beta$, which takes into account the number of boundaries. Accordingly, we
consider a slightly modified version of (\ref{t}) which better fits our purposes:
\begin{equation}\label{y}
e^{W(x,\lambda^{-1})} = \lambda^{\frac{\chi}{24} - 1 - \beta^{2}\frac{N}{2}}\exp\left(\sum_{g,h,n} 
\frac{\lambda^{2g - 2 + h + n}}{n!}\beta^{h}{\cal F}^{(g,h)}_{i_{1}\dots i_{n}}x^{i_{1}}\dots x^{i_{n}}\right)
\end{equation}
(\ref{y}) is obtained from (\ref{t}) suppressing all the open moduli parameters $u^{\alpha}$, rescaling $x^{i}\rightarrow \lambda x^{i}$ and 
considering the additional $\beta$-parameter dependence. 
The HAE for open strings of \cite{Wal1} are obtained as power expansion in $x^{i}$, $\lambda$ and  $\beta$ of (\ref{y}) 
\begin{equation}\label{z1}
\left(-\overline{\partial}_{\overline{i}} + \frac{1}{2}C^{jk}_{\overline{i}}\frac{\partial^{2}}{\partial x^{j}\partial x^{j}} 
+ G_{j\overline{i}}x^{j}\frac{\partial}{\partial\lambda^{-1}} - \beta\Delta_{\overline{i}}^{j}\frac{\partial}{\partial x^{j}} \right)
e^{W(x,\lambda^{-1})} = 0 .
\end{equation}

In \cite{Oog} it was shown that the above HAE (\ref{z1}) can be derived from the HAE of the closed theory
by means of a suitable change of variables 
\begin{eqnarray}\label{q}
x^{i} \rightarrow x^{i} + \beta\Delta^{i} \nonumber \\
\lambda^{-1} \rightarrow \lambda^{-1} - \beta\Delta 
\end{eqnarray}
with $\overline{\partial}_{\overline{i}}\Delta = \Delta_{\overline{i}}$ and 
$\overline{\partial}_{\overline{i}}\Delta^{i} = \Delta_{\overline{i}}^{i}$ such that 
$G_{i\overline{i}}\Delta^{i} = \Delta_{\overline{i}}$ and explicitly
\[
\Delta = g^{0\overline{0}}\int_{X}L_{CS}\wedge \overline{\Omega}_{0} \; \; \; \; g^{0\overline{0}} = 
\left(\int_{X}\Omega_{0} \wedge \overline{\Omega_{0}}\right)^{-1}
\]
\[
\Delta^{i} =g^{i\overline{j}} \left(\int_{X}L_{CS}\wedge d_{\overline{j}}\overline{\Omega}\right)_{x = 0} 
\; \; \; \; g^{i\overline{j}} = \left(\int_{X}d_{i}\Omega \wedge d_{\overline{j}}\overline{\Omega}\right)^{-1}_{x = 0} 
\]
where all the fields are on shell and $x=0$. Notice also that $\Delta$ and $\Delta^{i}$ have been computed starting from the 
antitopological theory. 
Finally the closed field does not appear because on shell it goes as $O(x^{2})$.
The shift (\ref{q}) allows to rewrite (\ref{z1}) in the same form as the master equation for purely closed strings
\begin{equation}\label{z2}
\left(-\overline{\partial}_{\overline{i}} + \frac{1}{2}C^{jk}_{\overline{i}}\frac{\partial^{2}}{\partial x^{j}\partial x^{j}} 
+ G_{j\overline{i}}x^{j}\frac{\partial}{\partial\lambda^{-1}}  \right)e^{W(x + \beta\Delta^{i},\lambda^{-1} - \beta\Delta)} = 0
\end{equation}
as follows from an easy application of the chain rule. Before going on let us mention that a refined shift was proposed in 
\cite{Wal2} in order to
have a detailed matching of the open and closed string amplitudes. 
The crucial point is that the change of variables proposed in \cite{Wal2} takes covariantly into account
the constraint over the amplitudes
\[
D_{i_{n}}{\cal F}^{(g,h)}_{i_{1}\dots i_{n-1}} = {\cal F}^{(g,h)}_{i_{1}\dots i_{n}}.
\]
%can be translated in an equation on the partition function much in the same way as (\ref{z1}) translates the H.A.E. . 
%The shift (\ref{q}) is able to reproduce the open H.A.E. starting from the closed version but misses the proper transformation of the constraint 
%to match the one for the open amplitudes. 
%The change of variables of \cite{Wal2} was proposed in order to take in account also this constraint.}. 
However, since we are interested in checking the fact that the integration over the open string modes produces a wave function satisfying the 
shifted closed HAEs, we can restrict ourselves to (\ref{q}). A more refined analysis of the boundary conditions would require
the calculation of the normalization factor $\caln$ in (\ref{yy}) corresponding to the rescaling in eq.(3.13) of \cite{Wal2}.

It is now possible to postulate that an analog shift for $x$ and $\lambda^{-1}$ in the path integral with the Kodaira-Spencer action 
(corresponding to the closed partition function) would allow to obtain the full path integral with the complete action. 

In order to reproduce the power expansion 
of (\ref{y}) from the target space field theory we have to set $x\to \lambda x$, so that any bulk operator insertion 
carries a weight $\lambda x$ as in (\ref{y}). 
To maintain our setting we translate (\ref{q}) into a shift for the product $\lambda x$ 
\begin{eqnarray}\label{z3}
\lambda x^{i} &\rightarrow & \lambda x^{i} + \lambda\beta\Delta^{i}  - \lambda^{2}\beta\Delta x^{i} + o(\lambda^{3},\beta^{2})\nonumber \\
\lambda^{-1} &\rightarrow & \lambda^{-1} - \beta\Delta 
\end{eqnarray}
of which we will keep only the lowest order term for the first line, discarding the $\lambda^{2}$ piece induced by the transformation of $\lambda$. 
From now on $\lambda x$ will be denoted simply  as $x$. We want to check that
\begin{equation}\label{v}
\int {\cal D}A e^{-S_{KS}(x^{i} + \lambda\beta\Delta^{i} + \dots,\lambda^{-1} - \beta\Delta;t,\overline{t};A)} 
\simeq \int {\cal D}A{\cal D}B{\cal D}b\dots e^{-S_{tot}(x,B_{0},\lambda^{-1};t,\overline{t};A,B,b,\dots)} 
\end{equation}
Let us consider (\ref{v}) at the tree level. Simply applying (\ref{z3}) to the Kodaira-Spencer action gives, 
at order $\beta$ and $\lambda^{-1}$, and redefining 
$S_{KS}$ in order to have the factor $\lambda^{-2}$ explicit,
\[
\frac{1}{\lambda^{2}}S_{KS}(x^{i} + \lambda\beta\Delta^{i} + \dots,\lambda^{-1} - \beta\Delta;t,\overline{t};A) 
= \frac{1}{\lambda^{2}}S_{KS}(x^{i},\lambda^{-1};t,\overline{t};A) - 
\]
\[
- \frac{\beta}{\lambda}\int_{M}[(A + x)(A + x)]'(\mu_{i})'\Delta^{i} - 
\frac{2\beta\Delta}{\lambda}S_{KS}(x^{i},\lambda^{-1};t,\overline{t};A) + O(\lambda^0,\beta^{2})
\]
Going at tree level the $O(\lambda^0,\beta^{2})$ are not taken into account; in addition the $A$ field should be 
taken on shell with respect to the Kodaira-Spencer equation in the shifted background, that is
\begin{equation}\label{u}
A \rightarrow A(x^{i} + \lambda\beta\Delta^{i} + \dots) = A(x) + \lambda\beta\Delta^{i}\partial_{i}A(x) + O(\lambda^{2},\beta^{2})
\end{equation}
Then, at order $\beta$, $\frac{1}{\lambda}$, the left side of (\ref{v}) is the exponential of
\[
\frac{1}{\lambda^{2}}S_{KS}(x^{i},\lambda^{-1};t,\overline{t};A(x)) 
- \frac{\beta}{\lambda}\int_{X}[(A(x) + x)(A(x) + x)]'(\mu_{i})'\Delta^{i} -
\]
\[
 - \frac{2\beta\Delta}{\lambda}S_{KS}(x^{i},\lambda^{-1};t,\overline{t};A(x)) +
\]
\begin{equation}\label{exp}
+ \frac{\beta}{\lambda}\int_{X}\Delta^{i}(\partial_{i}A(x))'\frac{1}{\partial}\overline{\partial}A(x) 
- [(A(x) + x)(A(x) + x)]'(\partial_{i}A(x))'\Delta^{i}  
\end{equation}
where the last line is actually zero because of the equations obeyed by $A(x)$, and the second line reduces to 
\[
-\frac{\beta\Delta}{3\lambda}[(A(x) + x)(A(x) + x)]'(A(x) + x)' = 
\]
\[
= -\frac{\beta\Delta}{\lambda}[(A(x) + x)(A(x) + x)(A(x) + x)]'\Omega_{0} 
\]
Remembering the expression (\ref{x}) we can substitute the value of $\Delta^{i}$ in (\ref{exp}) and get, for the second term 
in the first line of (\ref{exp}),
\begin{eqnarray}\label{s}
\frac{\beta}{\lambda}\int_{X}\Omega^{(1,2)}_{A = A(x)}\wedge (d_{i}\Omega)^{(2,1)}_{x = 0}\left(\int_{X}(d_{i}\Omega)^{(2,1)}_{x = 0} 
\wedge (d_{\overline{j}}\overline{\Omega})^{(1,2)}_{\overline{x} = 0}\right)^{-1}\cdot \nonumber \\
\cdot \int_{X}L_{CS}^{(2,1)}\mid_{B = B(u,x)}\wedge (d_{\overline{j}}\overline{\Omega})^{(1,2)}_{\overline{x} = 0}   
 = \frac{\beta}{\lambda}\int_{X}{\Omega}^{(1,2)}_{A = A(x)}\wedge L_{CS}^{(2,1)}\mid_{B = B(u,x)}
\end{eqnarray}
The last equality has been obtained using the Riemann bilinear relations:
\[ \int_{X}w\wedge\hat{w} = \sum_{a = 0}^{h_{2,1}}\int_{\delta_{a}}w\int_{\delta_{a + h_{2,1}}}\hat{w} 
- \int_{\delta_{a + h_{2,1}}}w\int_{\delta_{a}}\hat{w} 
\]
where $\delta_{a}$ is a base of 3-cycles on X.
First we express in this way the integrals containing $\Omega \wedge d_{i}\Omega$ and 
$ L_{CS}\wedge d_{\overline{j}}\overline{\Omega}$. 
Then we can define $X^{i}$ and $\overline{X}^{j}$ as three forms such that
\[
\left(\int_{X}d_{i}\Omega\wedge d_{\overline{j}}\overline{\Omega}\right)^{-1} \equiv \int_{X}X^{i}\wedge \overline{X}^{j} 
= \sum_{a = 0}^{h_{2,1}}\int_{\delta_{a}}X^{i}\int_{\delta_{a + h_{2,1}}}\overline{X}^{j} 
- \int_{\delta_{a + h_{2,1}}}X^{i}\int_{\delta_{a}}\overline{X}^{j}  
\]
respects the definition
\[
\sum_{\overline{j}}\left(\int_{X}d_{i}\Omega\wedge d_{\overline{j}}\overline{\Omega}\right)^{-1}
\int_{X}d_{k}\Omega \wedge d_{\overline{j}}\overline{\Omega} = \delta_{i,k}
\]
that is  
\[
\sum_{i}\int_{\delta_{a}}d_{i}\Omega\int_{\delta_{b}}X^{i} \equiv \delta_{a,b} \;\; \;\;\; \sum_{a = 0}^{2h_{2,1} +
 2}\int_{\delta_{a}}d_{i}\Omega\int_{\delta_{a}}X^{j}\equiv \delta_{i,j}
\]
and similarly with the barred quantities. Substituting these expressions in (\ref{s}) we obtain the result. 

Equivalently for the term in $\Delta$ in the second line of (\ref{exp}) we get 
\begin{equation}
\frac{\beta}{\lambda}\int_{X}{\Omega}^{(0,3)}_{A = A(x)}\wedge L_{CS}^{(3,0)}\mid_{B = B(u,x)}
\end{equation}
In order to reconstruct the full integral $\int_{X}{\Omega}_{A = A(x)}\wedge L_{CS}\mid_{B = B(u,x)}$
from the above equation the $(0,3)$ and $(1,2)$ components of $L_{CS}$ are still missing.
Notice however that they can be recovered by requiring CPT invariance. In particular,   
we modify (\ref{s}) as
\[
\frac{\beta}{\lambda}\int_{X}\left(\Omega^{(1,2)}_{A = A(x)} + \Omega^{(2,1)}_{A = A(x)}\right)
\wedge (d_{i}\Omega)^{(2,1)}_{x = 0}\cdot g^{i\overline{j}}\cdot
\]
\[
\cdot\int_{X}\left(L_{CS}^{(2,1)}\mid_{B = B(u,x)} + L_{CS}^{(1,2)}\mid_{B = B(u,x)}\right)
\wedge (d_{\overline{j}}\overline{\Omega})^{(1,2)}_{\overline{x} = 0} 
\]
where the extra term actually vanishes due to form degree reasons.   
This lead to an additional term
\[
 \frac{\beta}{\lambda}\int_{X}{\Omega}^{(2,1)}_{A = A(x)}\wedge L_{CS}^{(1,2)}\mid_{B = B(u,x)}
\]
An analogous modification has to be performed in order to obtain the 
$(0,3)$ component of $L_{CS}$. 

The geometrical counterpart of the above is as follows.
We know from the discussion of \cite{Wal1} that 
the coupling of the on-shell Chern-Simons action to $\Omega_0$ can be translated in mathematical terms to the pairing with the related 
normal function, $\nu$, dual to a suitable three-chain, $\Gamma$, such that
\[
\int_{X}\Omega_{0}\wedge L_{CS}\mid_{B=B(u,x)} = \int_{\Gamma}\Omega_{0} = \langle \Omega_0, \nu \rangle 
\]
and similarly for a $(2,1)$ form. Then it exists a lift of $\nu$ such that the coupling with a $(0,3)$ and $(1,2)$ forms are defined 
to be obtained by CPT invariance, that is complex conjugation of the corresponding $(0,3)$ and $(2,1)$ couplings. 

Summarizing we have shown that
\begin{eqnarray}\label{r}
\frac{1}{\lambda^{2}}S_{KS}(x^{i} + \beta\lambda\Delta^{i},\lambda^{-1} - \beta\Delta;t,\overline{t})\mid_{on \; shell} = \nonumber \\  
=\left(\frac{1}{\lambda^{2}}S_{KS}(x^{i},\lambda^{-1};t,\overline{t}) + \frac{\beta}{\lambda}\int_{X}\Omega \wedge L_{CS} 
- \Omega db 
%+  {\bar\psi}F_{B_{0}}^{\tilde{2},\tilde{0}}
\right)\mid_{on \; shell}
\end{eqnarray}
in the gauge $F_{B_{0}}^{\tilde{2},\tilde{0}}=0$.
Notice that the completion of the solution via CPT invariance obtained by adding the classical solutions of the anti-topological theory
is consistent with the fact that, in our gauge, the gauge fixing $F^{(2,0)}=0$ and the equation of motion $F^{(0,2)}=0$
of the topological theory are the same, up to a switch of role, as in the
{\it on shell} anti-topological one which is then manifestly CPT conjugate.

%Notice that it is the enforcing of the same equations of motion obeyed by $A(x)$ on both sides which explains the appearing 
%of the $b$ field and the 
%Lagrange multiplier ${\bar\psi}$. In particular for what concerns $b$ its role is to enforce for the right hand side of (\ref{r}) 
%the same equation obeyed by 
%$A(x)$ on the left. ${\bar\psi}$ instead enforces on the right the constraint $F_{B_{0}}^{\tilde{2},\tilde{0}} = 0$, which 
%correspond on the left to the equation of motion from the antitopological theory from which $B$ has came from, while the 
%actual equation of motion for the $B$ field on the right, $F_{B_{0}}^{\tilde{0},\tilde{2}} = 0$, corresponds to the previous 
%zero modes constrain on the left.  

\section{Conclusions}

In this paper we provided a target space string field theory formulation for open and closed B-model,
by giving a BV quantization of the holomorphic Chern-Simons theory with off shell gravity background.
This allowed us to design a target space interpretation of the coefficients in the HAE
with open moduli in general. 
In this paper we applied our formalism to reproduce the results of \cite{Oog} and interpret them as 
an open/closed string duality.
It would be interesting to study other explicit examples 
to refine the details of the scheme that we have been elaborating so far:
on the conifold the on shell results of \cite{K} could be useful.

Moreover, the target space formulas we obtained for the structure coefficients
of the HAE should complete the data needed to rephrase the latter as conditions of
background independence of the open B-model wave-function extending \cite{witten-ind}\cite{Wal2}.

The picture we provided in this paper seems to allow an extension to generalized complex geometries.
This should follow by the definition of an extended Chern-Simons functional
where the 3-form $\Omega$ gets promoted to the relevant pure spinor
as in \cite{luca}. Once this is done and the $b$ field promoted to a multiform, this would extend to open strings
the proposal in \cite{pestun} to generalized complex geometry of an analog of the Kodaira-Spencer theory.

\vspace{1 cm}

{\bf Acknowledgements} We thank Camillo Imbimbo, Sara Pasquetti, Emanuel Scheidegger, Johannes Walcher and Jie Yang for useful discussions.
%Moreover, we thank Andrei Losev for his trick.

\end{document}